\documentclass{icrc29}
\usepackage{graphicx,amssymb,amsmath,times}
\begin{document}
\title[Search for VHE Emission from GRBs with Milagro]{Search for Very High Energy Emission from Satellite-triggered GRBs with the Milagro Observatory}

\author[P.~M.~Saz Parkinson]{P.~M.~Saz Parkinson$^a$ for the Milagro Collaboration \\
	(a) Santa Cruz Institute for Particle Physics, University of California, 1156 High Street, Santa Cruz, CA 95064 }

\presenter{Presenter: P.~M.~ Saz Parkinson (pablo@scipp.ucsc.edu), \  
usa-saz-parkinson-P-abs1-og24-oral}

\maketitle

\begin{abstract}
The Milagro gamma-ray observatory employs a water Cherenkov detector to 
observe extensive air showers produced by high energy particles interacting 
in the Earth's atmosphere. Milagro has a wide field of view (2 sr) and high duty 
cycle ($>$ 90\%) making it an ideal all-sky monitor of the northern hemisphere in the 
100 GeV to 100 TeV energy range. More than 45 satellite-triggered gamma-ray bursts (GRBs) 
have occurred in the field of view of Milagro since January 2000, with the rate of bursts
increasing significantly with the launch of \emph{Swift}. We discuss 
the most recent results of a search for very high energy (VHE) emission from these GRBs.  
\end{abstract}

\section{Introduction}

The search for prompt VHE emission ($>$100 GeV) from GRBs is motivated by experimental 
observations and theoretical predictions, and its detection could
allow us to constrain GRB emission models. Although VHE emission from GRBs has not been conclusively 
demonstrated, there are several hints of emission at these high energies.
Milagrito, a prototype of Milagro, searched for emission coincident with 54 BATSE bursts
and reported evidence for emission above 650 GeV from GRB 970417a, with a (post-trials) probability of 
1.5$\times10^{-3}$ of being a background fluctuation~\cite{atkins00a,atkins03}. 
The HEGRA group reported evidence at the 3 sigma level for emission above 20 TeV from GRB 920925c~\cite{padilla98}. 
Follow-up observations above 250 GeV by the Whipple atmospheric Cherenkov telescope~\cite{connaughton97} failed to
find any high energy afterglow for 9 bursts studied, though the delay in slewing to observe these
bursts ranged from 2 minutes to almost an hour.
EGRET detected emission above 100 MeV from several bursts, including an 18 GeV photon from 
GRB 940217, over 90 minutes after the start of the burst~\cite{hurley94}, indicating both that the spectra of 
some GRBs extend to at least GeV energies and that this emission may be delayed~\cite{dingus95,dingus01}.
More recently, a second spectral component was discovered in GRB 941017~\cite{gonzalez03} which extended up to at 
least 200 MeV and decayed more slowly than the lower energy component.

While VHE emission from GRBs has been elusive, many GRB production models predict a fluence at TeV energies
comparable to that at MeV energies~\cite{dermer00,pilla98,zhang01}. This is because MeV 
emission from GRBs is likely synchrotron radiation produced by energetic electrons within the strong 
magnetic field of a jet with bulk Lorentz factors exceeding 100. In such an environment,
the inverse Compton mechanism for transferring energy from electrons to gamma rays is 
likely to produce a second higher energy component of GRB emission with fluence possibly
peaked at 1 TeV or above.  The relative strengths of the synchrotron and inverse Compton
emission depend on the environments of the particle acceleration and the gamma ray 
production.

Milagro\cite{atkins00b,atkins01} is a TeV gamma-ray detector, located at an altitude of 2630 m in
northern New Mexico, which uses the water Cherenkov technique to detect extensive air-showers 
produced by VHE gamma rays as they interact 
with the Earth's atmosphere. Its field of view is $\sim$2 sr and duty cycle $>$90\%. The effective area 
is a function of zenith angle and ranges from $\sim50$ m$^2$ at 100 GeV to $\sim10^5$ m$^2$ at 10 TeV. A 
sparse array of 175 4000-l water tanks, each with a PMT, was added in 2002. These ``outriggers,'' extend the 
physical area of Milagro to 40000 m$^2$, substantially increasing its sensitivity. The angular resolution is 
approximately $0.75^\circ$. The combination of large field of view and high duty cycle make Milagro the best 
instrument available for conducting a search for VHE emission from GRBs. 

Twenty-five satellite-triggered GRBs occurred within the field of view of Milagro between 
January 2000 and December 2001. No significant emission was detected from any of these bursts~\cite{atkins05}. 
Between January 2002 and mid-December 2004 only 11 well-localized GRBs were within $45^\circ$ of zenith at 
Milagro, due to the demise of BATSE. However, in the six months since the launch of 
\emph{Swift}~\cite{2004ApJ...611.1005G}, an additional 
11 bursts have fallen in Milagro's field of view~\footnote{For information about well-localized GRBs see J. Greiner's web 
page http://www.mpe.mpg.de/$\sim$jcg/grbgen.html}, several of them with measured 
redshift. Due to the absorption of high-energy gamma rays by the extragalactic background 
light, detections at VHE energies are only expected for redshifts less than $\sim$0.5. The degree of gamma-ray 
extinction from this effect is uncertain, because the amount of EBL is not well known. There are different models of the 
extinction~\cite{stecker98,dejager02,primack99}, which are similar in their general features because of the constraints from 
the available data. The most recent model now predicts a somewhat smaller absorption than was previously 
expected~\cite{primack04}, with an optical depth predicted to be roughly unity for 500 GeV (10 TeV) gamma rays from a 
redshift of 0.2 (0.05).

In addition to searching for VHE emission from satellite-localized GRBs, an independent 
real-time search for VHE bursts in the Milagro data has been conducted for many different durations~\cite{noyes}. 

\section{The GRB sample}

Table~\ref{grb_table} shows a summary of the sample of satellite-triggered GRBs within the field of 
view (up to zenith angles of $45^{\circ}$) of Milagro between January 2002 and June 2005. 
The first column lists the GRB name, following the usual convention (UTC date YYMMDD). The second column gives the 
instrument that first reported the burst. The third column gives the trigger time (UTC second of the day).
Column four gives the coordinates, right ascension (RA) and declination (Dec.), in degrees, of the burst. 
The fifth column gives the duration of the burst. Although this duration is derived from observations made at much 
lower energies than Milagro detects, EGRET showed that the T90 duration obtained in the keV regime is relevant at 
higher energies too. Four GRBs observed by EGRET were among the five brightest 
BATSE bursts, and the significance of the EGRET detections in the T90 interval ranged 
from 6 to 12 sigma, leading to the speculation that all GRBs might have high energy emission 
during their respective T90 time intervals, and EGRET simply did not have the sensitivity to detect most of them~\cite{dingus01}.
Column six of Table~\ref{grb_table} lists the zenith angle of the burst at Milagro, in degrees. We include only 
bursts for which the zenith angle was less than $45^{\circ}$ since the effective area of Milagro at zenith angles greater 
than $45^{\circ}$ becomes small in the energy range where we expect GRB emission to be detectable (e.g. $<$ 1 TeV). 
Column 7 gives the redshift (if measured) of the burst. 

\begin{table}   
\begin{center}
\begin{tabular}{llllllllc} \hline \hline

GRB & 
Instrument &
UTC &
RA,Dec. & 
T90/Dur. & 
$\theta$ & 
z & 
Li-Ma &
99\% UL fluence \\
 &
 &
 &
(deg.) &
(s) &
(deg.) &
 &
$\sigma$ &
(erg cm$^2$) \\

\hline \hline

020625b	& HETE 		& 41149.3 & 310.9,+7.1 	& 125  	&  38.1 & ... 	& 1.4  & 5.7e-6 \\
021104 	& HETE 		& 25262.9 & 58.5,+38.0 	& 19.7 	&  13.3 & ... 	& 0.9  & 7.5e-7 \\
021112 	& HETE 		& 12495.9 & 39.3,+48.9  & 7.1	&  33.6 & ... 	& -0.1 & 9.4e-7 \\
021113 	& HETE 		& 23936.9 & 23.5,+40.5  & 20	&  17.7 & ... 	& 0.1  & 6.4e-7 \\
021211 	& HETE 		& 40714.0 & 122.3,+6.7  & 6 	&  34.8 & 1.01	& 2.0  & 1.7e-06$^*$ \\
030413 	& IPN 		& 27277.0 & 198.6,+62.4	& 15   	&  27.1	& ... 	& 0.8  & 1.0e-6 \\
030823 	& HETE 		& 31960.6 & 322.7,+22.0 & 56	&  33.4 & ... 	& 1.0  & 2.8e-6 \\
031026 	& HETE 		& 20143.3 & 49.7,+28.4  & 114.2 &  33.0 & ... 	& 0.7  & 3.8e-6 \\
031220 	& HETE 		& 12596.7 & 69.9,+7.4  	& 23.7 	&  43.4 & ... 	& 0.2  & 4.0e-6 \\
040924 	& HETE 		& 42731.4 & 31.6,+16.0  & 0.6  	&  43.3 &0.859	& -0.6 & 1.5e-06$^*$ \\
041211 	& HETE 		& 41507.0 & 101.0,+20.3	& 30.2 	&  42.8 & ... 	& 0.9  & 4.8e-6 \\
041219 	& INTEGRAL 	& 6400.0  & 6.1,+62.8  	& 520  	&  26.9 & ... 	& 1.7  & 5.8e-6 \\
050124 	& Swift 	& 41403.0 & 192.9,+13.0	& 4  	&  23.0 & ...	& -0.8 & 3.0e-7 \\
050319 	& Swift 	& 34278.4 & 154.2,+43.5	& 15  	&  45.1 & 3.24	& 0.6  & 4.4e-06$^*$\\
050402 	& Swift 	& 22194.6 & 136.5,+16.6	& 8  	&  40.4 & ...	& 0.6  & 2.1e-6 \\
050412 	& Swift 	& 20642.9 & 181.1,-1.3 	& 26  	&  37.1 & ...	& -0.6 & 1.7e-6 \\
050502 	& INTEGRAL 	& 8057.7  & 202.4,+42.7	& 20  	&  42.7 & 3.793	& 0.6  & 3.8e-06$^*$ \\
050504 	& INTEGRAL 	& 28859.1 & 201.0,+40.7	& 80  	&  27.6 & ...	& -0.8 & 1.3e-6 \\
050505 	& Swift 	& 84141.1 & 141.8,+30.3	& 60  	&  28.9 & 4.3	& 1.2  & 2.3e-06$^*$\\
050509b	& Swift 	& 14419.2 & 189.1,+29.0	& 0.03  &  10.0 & 0.225 & -0.9 & 9.2e-08$^*$ \\
050522 	& INTEGRAL 	& 21621.0 & 200.1,+24.8	& 15 	&  22.8 & ...	& -0.6 & 5.1e-7 \\
050607 	& Swift 	& 33082.7 & 300.2,+9.1 	& 26.5  &  29.3 & ...	& -0.9 & 8.9e-7 \\

\hline
\end{tabular}
\end{center}
\caption{\label{grb_table} GRBs in the field of view of Milagro in 2002--2005.
The upper limits assume the burst was nearby (z=0). Those with a (*) next to them have a 
measured redshift, making this assumption invalid (See text).}
\end{table}

\section{Data Analysis and Results}

A search for an excess of events above the expected background was
carried out for the 22 bursts listed in Table~\ref{grb_table}. The total number of events within a 
circular bin of radius $1.6^{\circ}$ at the location of the burst was summed for the duration of 
the burst and the number of background events was extimated by characterizing the angular 
distribution of the background using two hours of data surrounding the burst~\cite{atkins03}. The 
significance of the excess (or deficit) for each burst was evaluated using equation 17 of~\cite{lima} 
and is shown in column 8 of Table~\ref{grb_table}.

No significant excess was found from any burst in the sample. The 99\% confidence upper limit on 
the number of signal events detected given the observed events and the predicted background is 
computed for each of the bursts following the method described by~\cite{helene}, except for 
those bursts where the number of events is small ($<$ 10), where we use the prescription by~\cite{feldman}. 
Finally, we convert the upper limit on the counts into an upper limit on the fluence (in the 0.25 to 25 TeV range) 
by using knowledge of the effective area of Milagro, and assuming a differential power-law 
photon spectrum of the form $dN/dE=KE^{-2.4}$ photons/TeV/m$^2$. The power-law index of 2.4 was 
chosen as a conservative value for the spectrum. The upper limits listed in column 9 of Table~\ref{grb_table} assume 
the bursts occured nearby (z=0), ignoring the effects of the EBL absorption. This assumption is invalid for 
the six bursts with measured redshifts. For those with redshifts $>1$, all TeV emission would be absorbed. 
Only two bursts in our sample have redshifts less than 1. For GRB 040924, taking into account the absorption, 
the upper limits on the fluence are 1.6$\times10^{-3}$ erg cm$^{-2}$ and 3.1$\times10^{-3}$ erg cm$^{-2}$, using the
absorption models of Primack et al.~\cite{primack04} and Stecker et al.~\cite{stecker98} respectively. We describe
the most interesting GRB of our sample in the next section.

\subsection{GRB 050509b}

GRB 050509b~\cite{hurkett} is so far only the second \emph{short/hard} burst detected by
\emph{Swift}. It had a reported duration of 30 ms and a relatively low fluence of 
2.3$\times10^{-8}$ erg cm$^{-2}$ in the 15--350 keV range~\cite{barthelmy}. This bursts 
represents the first clear detection of an afterglow from a short burst~\cite{kennea}.
Although Milagro detected no emission from this burst~\cite{saz}, the very favorable zenith 
angle ($10^\circ$) and relatively low redshift of 0.225~\cite{bloom}\footnote{Some observers still
question whether this is the true redshift of the burst~\cite{castro-tirado}.} provide the opportunity
to set interesting upper limits for TeV emission from this burst. 
Assuming a differential photon spectral index of -2.4, the derived 99\% upper limit on 
E$^2$dN/dE at 2.5 TeV is 5.4$\times10^{-8}$ erg cm$^{-2}$, assuming 
no EBL absorption. Taking into account the attenuation of TeV photons expected at a redshift
of 0.225, the 99\% upper limits for E$^2$dN/dE are 5.5$\times 10^{-7}$ erg cm$^{-2}$ at 150 GeV, using 
the model of Stecker et al.~\cite{stecker98}, and $2.0 \times 10^{-7}$ erg cm$^{-2}$ at 300 GeV, using 
the model of Primack et al.~\cite{primack04}. The energies quoted represent the median energy of the 
events that would be detected from a power-law spectrum with index -2.4 convolved with each absorption model.

\section{Acknowledgements}

Many people helped bring Milagro to fruition. In particular, we
acknowledge the efforts of Scott DeLay and Michael Schneider. 
This work has been supported by the National Science Foundation (under grants 
PHY-0075326, 
-0096256, 
-0097315, 
-0206656, 
-0245143, 
-0245234, 
-0302000, 
and
ATM-0002744) 
the US Department of Energy (Office of High-Energy Physics and 
Office of Nuclear Physics), Los Alamos National Laboratory, the University of
California, and the Institute of Geophysics and Planetary Physics.

\end{document}